\documentclass[onecolumn,11pt]{IEEEtran}

\usepackage{url}
\usepackage{underscore}
\usepackage{amsmath,amssymb}
\usepackage{mathtools} 
\usepackage{mleftright}  
\usepackage{bm,bbm} 
\usepackage{siunitx}

\usepackage[dvipsnames]{xcolor}
\usepackage{graphicx}
\usepackage{float}
\usepackage{wrapfig}

\usepackage{cite}

\usepackage[color=red!30,colorinlistoftodos]{todonotes}

\def\mc{\mathcal}

\def\mbb{\mathbb}

\newcommand{\R}{\mathbb{R}}

\DeclareMathOperator*{\minimize}{\mathrm{minimize}}
\DeclarePairedDelimiter\parens{\lparen}{\rparen}  

\DeclarePairedDelimiter\norm{\lVert}{\rVert}

\NewDocumentCommand{\expect}{ e{_} s o >{\SplitArgument{1}{|}}m }{%
  \operatorname{\mathbb{E}}
  \IfValueT{#1}{{\!}_{#1}}
  \IfBooleanTF{#2}{
    \expectarg*{\expectvar#4}%
  }{
    \IfNoValueTF{#3}{
      \expectarg{\expectvar#4}%
    }{
      \expectarg[#3]{\expectvar#4}%
    }%
  }%
}
\NewDocumentCommand{\expectvar}{mm}{%
  #1\IfValueT{#2}{\nonscript\;\delimsize\vert\nonscript\;#2}%
}
\DeclarePairedDelimiterX{\expectarg}[1]{[}{]}{#1}

\NewDocumentCommand{\prob}{ e{_} s o >{\SplitArgument{1}{|}}m }{%
  \operatorname{\mathbb{P}}
  \IfValueT{#1}{{\!}_{#1}}
  \IfBooleanTF{#2}{
    \probarg*{\probvar#4}%
  }{
    \IfNoValueTF{#3}{
      \probarg{\probvar#4}%
    }{
      \probarg[#3]{\probvar#4}%
    }%
  }%
}
\NewDocumentCommand{\probvar}{mm}{%
  #1\IfValueT{#2}{\nonscript\;\delimsize\vert\nonscript\;#2}%
}
\DeclarePairedDelimiterX{\probarg}[1]{(}{)}{#1}

\newcommand{\condon}{\,\ifnum\currentgrouptype=16 \middle\fi|\,} 

\newcommand{\fdiv}[3]{D_{#1}\parens*{#2 \mid\mid #3}} 
\newcommand{\Egam}{\mathsf{E}_{\gamma}}
\newcommand{\Ediv}[2]{\Egam\parens*{#1 \mid\mid #2}} 

\newcommand{\kldiv}[2]{\fdiv{\mathrm{KL}}{#1}{#2}} 

\newcommand{\Pfa}{P_{\mathrm{FA}}}
\newcommand{\Pmd}{P_{\mathrm{MD}}}

\newcommand{\out}{y}
\newcommand{\rout}{Y}
\newcommand{\aout}{\mc{Y}} 
\newcommand{\plrv}{L} 

\newcommand{\dpeps}{\varepsilon}
\newcommand{\dpdel}{\delta}

\graphicspath {{Figures/}}

\begin{document}

\title{An information theorist's tour of differential privacy}

\author{%
Anand D.~Sarwate, 
Flavio P. Calmon, 
Oliver Kosut, 
Lalitha Sankar 
%
\thanks{Icons/emoji in images courtesty of OpenMoji.org.}}%


\maketitle

\begin{abstract}
Since being proposed in 2006, differential privacy has become a standard method for quantifying certain risks in publishing or sharing analyses of sensitive data. At its heart, differential privacy measures risk in terms of the differences between probability distributions,  which is a central topic in information theory. A differentially private algorithm is a channel between the underlying data and the output of the analysis. Seen in this way, the guarantees made by differential privacy can be understood in terms of properties of this channel. In this article we examine a few of the key connections between information theory and the formulation/application of differential privacy, giving an ``operational significance'' for relevant information measures.
\end{abstract}


\section{Introduction}

\IEEEPARstart{D}{}ifferential privacy~\cite{DworkMNS:06sensitivity} is a framework for quantifying privacy risk as a property of an algorithm/computation operating on sensitive data. Privacy comes from randomizing the algorithm so that its output is a random variable whose distribution depends on the original data. This means that a differentially private algorithm acts as a conditional distribution, or \emph{channel}, between the data and its output. Privacy properties of the algorithm are therefore properties of this channel.

In the nearly two decades since the original definition was proposed~\cite{DworkMNS:06sensitivity}, differential privacy has been adopted in many different fields, most significantly in computer science and statistics. The most prominent deployment of differential privacy to date has been in the 2020 US Decennial Census~\cite{Abowd:18censusdp,Abowd19topdown}, which used differential privacy for releasing tabulations. There have also been several deployments by industry~\cite{ErlingssonPK:14rappor,DingKY:17microsoft,Apple:20dp,Near:18uber}.

In this article we will look at some key topics in differential privacy while ``wearing an information theory hat.'' There have been many surveys of differential privacy over the years~\cite{Dwork:08survey,SarwateC:13survey,JiLE:14survey,DworkRoth,DworkSSU:17exposed,PejoD:20sok,HsuMBSC:2021bits,Kim+21:privacy,OuadrhiriA:22survey,ZhaoChen:22survey,Li+:23survey,ProkhorenkovC:23survey,Pan+:24survey,Yang+:24survey}, including a previous article in this magazine by Hsu et al.~\cite{HsuMBSC:2021bits} which compared different metrics for privacy leakage. Here we will focus on differential privacy and its connections to hypothesis testing, generalized divergences, and channel optimization.

\section{From hypothesis testing to privacy measures}
\label{sec:hyp}


One way to understand differential privacy's guarantee is that the additional (``differential'') risk of a secret being revealed is bounded. The definition of differential privacy is equivalent to a statement about hypothesis testing, a connection which has been explored by several authors, most recently under the name ``$f$-differential privacy''~\cite{WassermanZ:10framework,KairouzOV:17composition,DongRS:21gaussiandp}. Here ``$f$'' refers to the shape of the error tradeoff curve in the hypothesis test. 

\subsection{Privacy risk and error tradeoffs}

\begin{figure}[t]
    \centering
    \includegraphics[width=0.7\columnwidth]{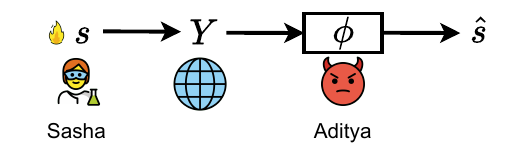}
    \caption{Guessing a single secret bit from leaked information.}
    \label{fig:onebitsecret}
\end{figure}

As a warmup, consider the example illustrated in Figure \ref{fig:onebitsecret}. Sasha has a binary secret $s \in \{0,1\}$ and information in the form of a random variable $\rout \in \R$ is revealed to the world. An adversary Aditya wishes to use $\rout$ to determine the value of $s$. For $\rout$ to reveal information about $s$ it must have a different distribution depending on whether $s = 0$ or $s = 1$. Let $P_{\rout|0}(\out)$ and $P_{\rout|1}(\out)$ denote these two conditional distributions. Aditya's problem is to find a decision rule using a realization $\rout = \out$ for the following binary hypothesis test:
    \begin{align}
        \mathcal{H}_0 &\colon \rout \sim P_{\rout|0}(\out) \\
        \mathcal{H}_1 &\colon \rout \sim P_{\rout|1}(\out).
    \end{align}
In a binary hypothesis test there are two types of error: false alarm $\Pfa$ (Type I) and missed detection $\Pmd$ (Type II). The Neyman-Pearson Lemma tells us that a randomized likelihood ratio test yields the optimal tradeoff between these two types of error. The privacy risk to Sasha from Aditya having $\rout$ is \emph{characterized by this tradeoff curve}.

We say that the pair $(P_{\rout|1},P_{\rout|0})$ guarantees $(\dpeps, \dpdel)$-differential privacy ($(\dpeps, \dpdel)$-DP) if their error tradeoff curve lies above a piecewise linear function~\cite{WassermanZ:10framework,KairouzOV:17composition} defined by
    \begin{align}
    \Pfa + e^{\dpeps} \Pmd &\ge 1 - \dpdel \label{eq:DPviatradeoff1}  \\
    e^{\dpeps} \Pfa + \Pmd &\ge 1 - \dpdel.\label{eq:DPviatradeoff2}
    \end{align}
This function is illustrated in Figure \ref{fig:dptradeoffs}. Many works refer to $(\dpeps, \dpdel)$-DP as ``approximate'' differential privacy. By taking $\delta = 0$ we get the definition of ``pure'' or $\dpeps$-differential privacy ($\dpeps$-DP).
    
\begin{figure}[tbh]
    \centering
    \includegraphics[width=0.4\columnwidth]{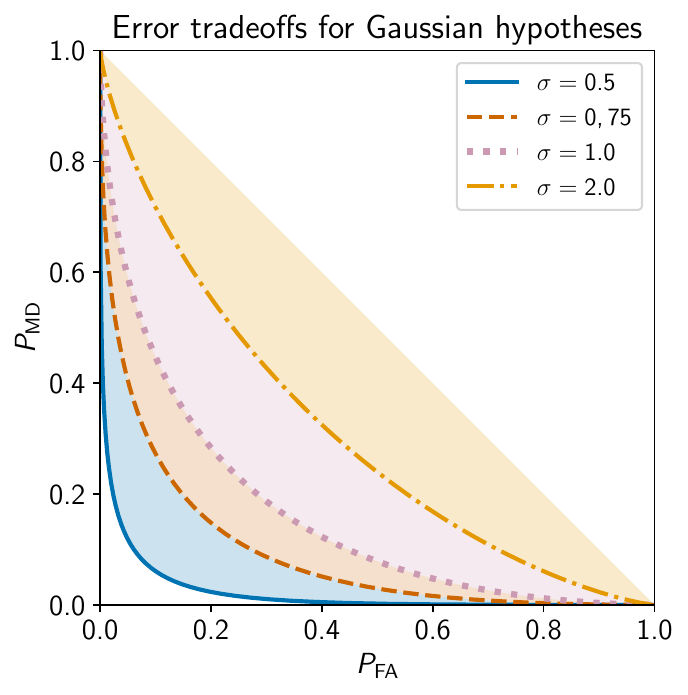}%
    \includegraphics[width=0.4\columnwidth]{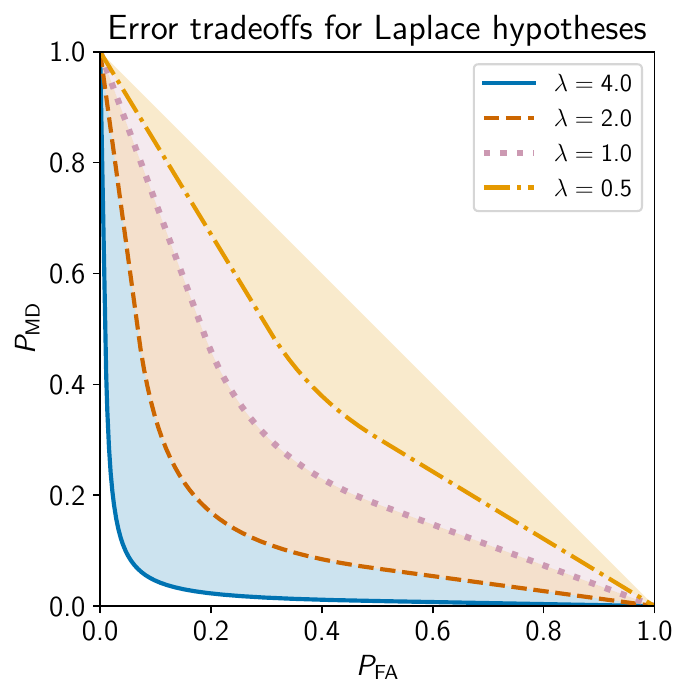}%
    \caption{Error tradeoffs for additive Gaussian and Laplace noise. For the Gaussian (left), $\rout \sim \mc{N}(0,\sigma^2)$ when $s = 0$ and $\rout \sim \mc{N}(1,\sigma^2)$ for $s = 1$. As $\sigma$ increases the error tradeoffs become closer and closer to the diagonal: more noise means more privacy since Aditya's test is harder. For the Laplace (right), $\rout$ has the distribution $\frac{\lambda}{2} \exp(- \lambda | \out - s |)$. As $\lambda$ increases the distribution becomes more concentrated and offers less privacy. The shapes of the tradeoffs differ depending on the distribution the noise.} 
    \label{fig:bhtcompare}
\end{figure}

In privacy mechanism design, Sasha gets to \emph{select the channel} (the distributions $P_{\rout|0}$ and $P_{\rout|1}$) and can use that freedom to make Aditya's hypothesis test more difficult.  Differential privacy corresponds to a lower bound on the error tradeoff curve $\Pmd(\Pfa)$, raising the natural question: can Sasha design $(P_{\rout|1},P_{\rout|0})$ to match this lower bound? A simple strategy for Sasha could be to add noise to $s$: Figure \ref{fig:bhtcompare} shows some possible choices of tradeoff curves between $\Pfa$ and $\Pmd$ corresponding to Sasha adding zero-mean Gaussian or Laplace noise. Of course, to make Aditya's job maximally difficult, Sasha could choose $\rout$ to be independent of $s$. The central tension in differential privacy is that Sasha wants to produce an $\rout$ which is \emph{useful} (and thus depends on $s$). She must therefore manage a \emph{privacy-utility tradeoff}.

\subsection{Understanding the privacy loss random variable}

Moving away from the specific example but sticking with a single secret bit $s$, we can consider $\rout$ that takes values in a general space $\aout$ and with $P_{\rout|0}$ and $P_{\rout|1}$ having density functions on $\aout$. 
Aditya's optimal tests are threshold tests on the log likelihood ratio, denoted by 
    \begin{align}
    \plrv(\rout) = \log \frac{ P_{\rout|1}(\rout) }{ P_{\rout|0}(\rout) }.
    \label{eq:plrv}
    \end{align}
To understand privacy risk, we need to understand the properties of $\plrv(\rout)$, which is a random variable whose randomness comes from $\rout$. The distribution of $\rout$ will differ depending on whether $s = 0$ or $s = 1$. The variable $\plrv$ is called the \emph{privacy loss random variable (PLRV)}. The distribution of $\plrv$ is sometimes called \emph{the privacy loss distribution (PLD)}~\cite{sommer2019privacy,zhu2022optimal}.

A starting point for understanding $\plrv$ is to look at its mean:
    \begin{align}
    \expect{ \plrv } = \begin{cases} 
        \kldiv{P_{\rout|1}}{P_{\rout|0}} & \rout \sim P_{\rout|1} \\ 
        -\kldiv{P_{\rout|0}}{P_{\rout|1}} & \rout \sim P_{\rout|0}
    \end{cases}.\label{kl-divergence}
    \end{align}
The expected PLRV is a Kullback-Leibler (KL) divergence $\kldiv{\cdot}{\cdot}$. This divergence is only one of many ways to quantify how different $P_{\rout|1}$ and $P_{\rout|0}$ are: in Section \ref{sec:divergences} we will look at the distribution of $\plrv$ by using other divergences.

\begin{figure}[tbh]
    \centering
    \includegraphics[width=0.4\columnwidth]{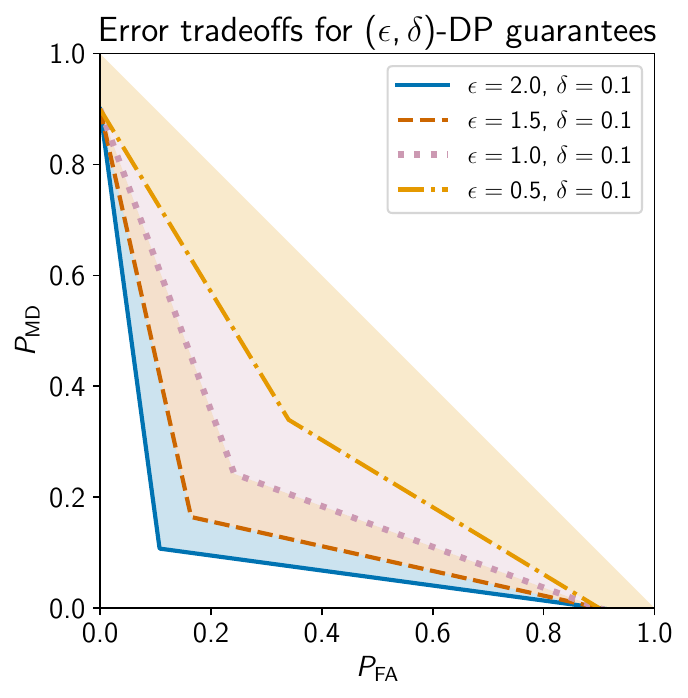}%
    \includegraphics[width=0.4\columnwidth]{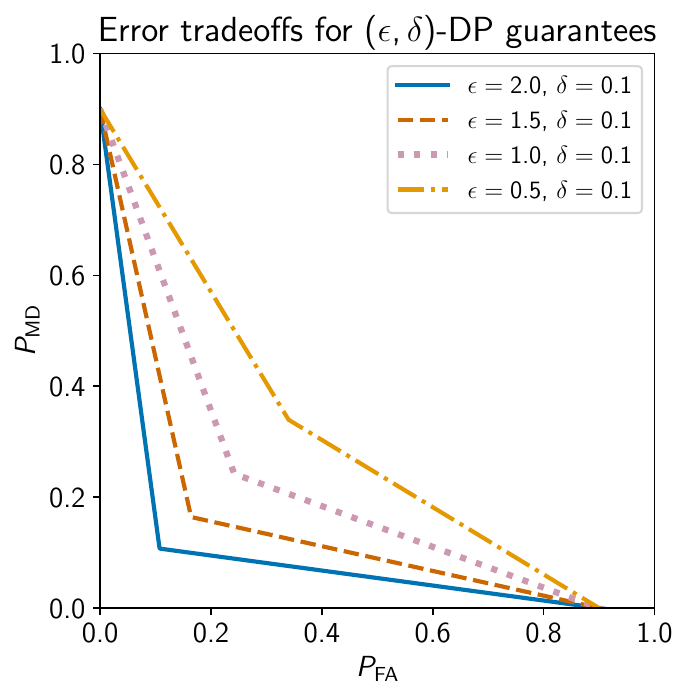}%
    \caption{(Left) error tradeoffs $(\dpeps,\dpdel)$-differential privacy as shown defined in \eqref{eq:DPviatradeoff1}--\eqref{eq:DPviatradeoff2}. (Right) Comparison of the tradeoff curve shapes for $(\dpeps,\dpdel)$-DP, additive Gaussian noise, and additive Laplace noise.} 
    \label{fig:dptradeoffs}
\end{figure}

Properties of the the PLRV imply the differential privacy definition we saw before. For example, if $\plrv$ has finite support so that $\plrv \in [\dpeps,-\dpeps]$ almost surely, then the distributions $(P_{\rout|1},P_{\rout|0})$ guarantee $\dpeps$-DP. If $\mathbb{P}( |\plrv| < \dpeps ) \ge 1 - \dpdel$, then they guarantee $(\dpeps, \dpdel)$-DP \footnote{However, this condition is not equivalent to the definition of $(\dpeps, \dpdel)$-DP. That is, a pair of distributions can guarantee $(\dpeps, \dpdel)$-DP even while $\mathbb{P}( |\plrv| < \dpeps ) < 1 - \dpdel$. The proper definition is in \eqref{eq:DPviatradeoff1}--\eqref{eq:DPviatradeoff2}.}.

Figure \ref{fig:dptradeoffs} illustrates lower bounds corresponding to different values of $(\dpeps,\dpdel)$ in \eqref{eq:DPviatradeoff1}--\eqref{eq:DPviatradeoff2} as well as a comparison between the $(\dpeps, \dpdel)$-DP tradeoff and the Gaussian and Laplace tradeoffs we saw in Figure \ref{fig:bhtcompare}. This comparison demonstrates that any lower bound on the error tradeoff can define a privacy metric. This way of thinking follows the framework of Blackwell's comparison of experiments~\cite{Blackwell:1950comparison,Blackwell:51comparison,Blackwell:53equivalent,Raginsky11:experiments}. The recently proposed $f$-differential privacy framework~\cite{DongRS:21gaussiandp} also takes this perspective: if there is a function $f(\cdot)$ such that for the given mechanism, $f(P_{\mathrm{FA}}) \le P_{\mathrm{MD}}(P_{\mathrm{FA}})$, then the mechanism guarantees $f$-differential privacy.

Gaussian DP (GDP)~\cite{DongRS:21gaussiandp} chooses $f$ to be the error tradeoff curve $\Pmd(\Pfa)$ for testing between two Gaussians $\mc{N}(0,1)$ and $\mc{N}(\mu,1)$ for $\mu > 0$. This replaces the piecewise linear lower bound in \eqref{eq:DPviatradeoff1} and \eqref{eq:DPviatradeoff2} parameterized by $(\dpeps,\dpdel)$ with a curve parameterized by $\mu$. Analyzing channels in the GDP framework allows us to leverage our extensive knowledge of the Gaussian distribution.



\subsection{Beyond a single secret bit}

\begin{figure}[tbh]
    \centering
    \includegraphics[width=0.7\columnwidth]{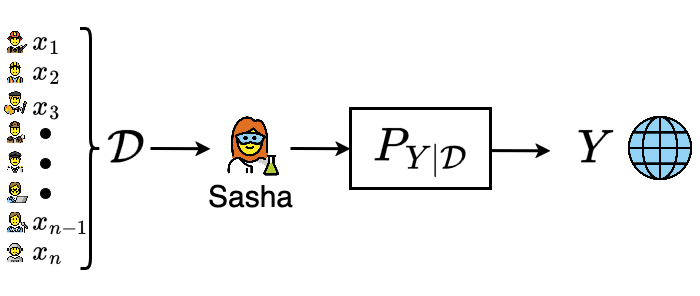}
    \caption{Differentially private channel mapping a database of individuals to an output $\rout$. Different values of $\mc{D}$ may induce different distributions on $\rout$.} 
    \label{fig:dpreveal}
\end{figure}

Differential privacy is about more than keeping a single bit secret. In the ``global'' or ``central'' model of differential privacy, we assume that Sasha has sensitive data from many individuals, as shown in Figure \ref{fig:dpreveal}. A single individual's data/record is an element $x$ taking values in a set $\mc{X}$. Sasha how has a dataset $\mc{D} = (x_1, x_2, \ldots, x_n)$ from $n$ individuals and wishes to release a random variable $\rout \in \mc{Y}$ whose distribution depends on $\mc{D}$. 
In the general case we want Aditya to have uncertainty about whether any particular $x_i$ is in $\mc{D}$ or not. We say a dataset $\mc{D}'$ with $n$ records is a neighbor of $\mc{D}$ if $|\mc{D} \cap \mc{D}'| = n-1$, meaning they differ in a single data record. Sasha wants Aditya to have high uncertainty in distinguishing \emph{any neighboring pair} $(\mc{D},\mc{D}')$. We will use $\mc{D} \sim \mc{D}'$ to denote that the pair $(\mc{D},\mc{D}')$ are neighboring. Instead of a single bit $s = 0$ or $s = 1$, Sasha now needs to design a mechanism that guarantees the same hypothesis testing tradeoffs \emph{simultaneously} for all pairs $\mc{D} \sim \mc{D}'$.

Changing notation slightly, Sasha's mechanism is now defined by a family of distributions $\{ P_{\rout | \mc{D} } \}$ from $\mc{X}^n \to \mc{Y}$. This mechanism guarantees $(\dpeps,\dpdel)$-differential privacy if for any $\out$ and any neighboring pair $(\mc{D},\mc{D}')$ the error tradeoff curve for a test between $\mc{D}$ and $\mc{D}'$ satisfies \eqref{eq:DPviatradeoff1}-\eqref{eq:DPviatradeoff2}. An easier-to-check sufficient condition is that for any measurable $T \subseteq \mc{Y}$ and neighboring pair $(\mc{D},\mc{D}')$ we have
   \begin{align}
   \mbb{P}_{P_{\rout | \mc{D}}}(\mc{T}) 
   \le e^{\dpeps} 
   \mbb{P}_{P_{\rout | \mc{D}'}}(\mc{T}) 
   + \delta.
    \label{eq:def:dp}
    \end{align}
This is more complex than our previous example of hiding a single bit because now the guarantee has to hold \emph{simultaneously} over all neighboring pairs $(\mc{D},\mc{D}')$. In many cases the analysis can be simplified because we can find a worst-case pair to analyze. It is important to note that \emph{any privacy mechanism $\{ P_{\rout | \mc{D} } \}$ achieves a range of $(\dpeps, \delta)$ values}. Nevertheless, in practice, it is custom to report the value of $\dpeps$ for a small $\delta$, usually for $\dpdel \ll 1/n$.\footnote{A mechanism which releases a single record chosen uniformly at random would guarantee $\dpdel = 1/n$.}

The PLRV for this more complex case can be defined for any pair of databases $\mc{D},\mc{D}'$ as
    \begin{align}
    \plrv_{\mc{D},\mc{D}'} = \log \frac{dP_{\rout|\mc{D}}}{dP_{\rout|\mc{D}'}}( \rout ).
    \label{eq:plrv2}
    \end{align}
and the distribution of $\plrv_{\mc{D},\mc{D}'}$ is the PLD for the pair $(\mc{D},\mc{D}')$. In the next section, we will see how this PLRV can tell us whether the mechanism satisfies $(\dpeps, \dpdel)$-DP. In fact, essentially \emph{any} meaningful measure of information can be derived from this variable.

\begin{figure}[tbh]
    \centering
    \includegraphics[width=0.7\columnwidth]{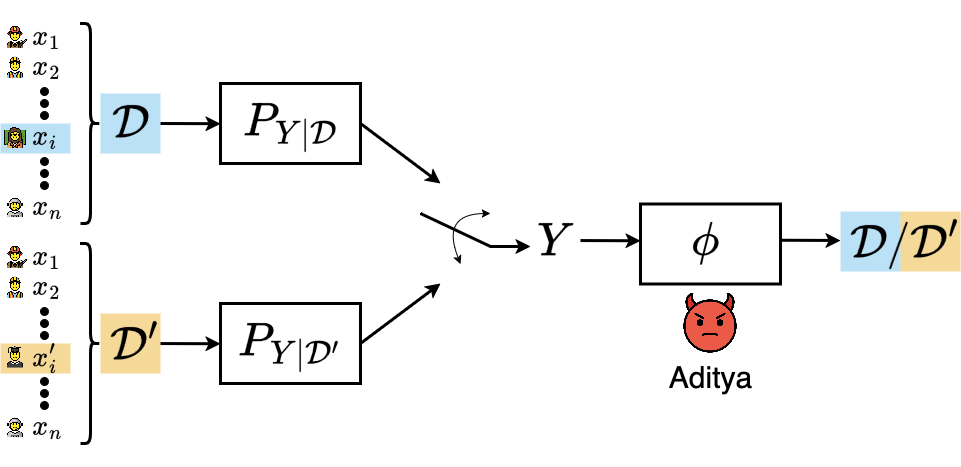}
    \caption{The adversary's inference problem of deciding between which of two neighboring databases generated the revealed observation $\rout$.} 
    \label{fig:dptest}
\end{figure}

\subsection{Comparison to earlier privacy measures}

Prior to the development of differential privacy, many approaches to privacy were designed for tabular data. Algorithms for ``anonymizing'' data for public release would manipulate data values through operations such as quantization (e.g. reporting ranges instead of values)~\cite{Sweeney:02kanon,MachanavajjhalaGKV:07diversity,LiLV:10tcloseness}. In these models, privacy was quantified as a property of the released data itself and not the algorithm which produced it. Approaches which used  randomization~\cite{AgrawalS:00mining,EvfimievskiGS:03breaches} did not quantify the privacy protection operationally in terms of hypothesis testing. By contrast, the privacy risk in differential privacy is quantified as a property of the algorithm (the channel) independent of the data. Moreover, the risk has an operational interpretation in terms of the family of hypothesis tests induced by the channel.

\section{Divergences and privacy definitions}
\label{sec:divergences}


We saw before that the expected value of the PRLV in \eqref{eq:plrv2} is a KL-divergence. Since divergences measure the discrepancy between two distributions, it is natural to ask if the differential privacy risk corresponds to some other divergence~\cite{Renyi61:measures,csiszar1963informationstheoretische,csiszar1967information,csiszar1967topological,morimoto1963markov,ali1966general,CsiszarS:04fnt}. Mironov~\cite{Mironov:17renyi} defined a model called R\'{e}nyi differential privacy (RDP) based on the R\'{e}nyi divergence~\cite{Renyi61:measures}, which allows for easier bounding and tracking of privacy loss across many iterations. This has proven to be useful in approximating the privacy loss for iterative algorithms such as noisy gradient descent used for training machine learning models~\cite{ACGMMTZ16}. It turns out that $(\dpeps,\dpdel)$-differential privacy can be equivalently formulated as an $f$-divergence~\cite{csiszar1967information}.\footnote{It is important to note that the ``$f$'' in $f$-differential privacy refers to a different function than the ``$f$'' in $f$-divergence.}

\subsection{$f$-divergences}

\begin{figure}
    \centering
    \includegraphics[width=0.7\columnwidth]{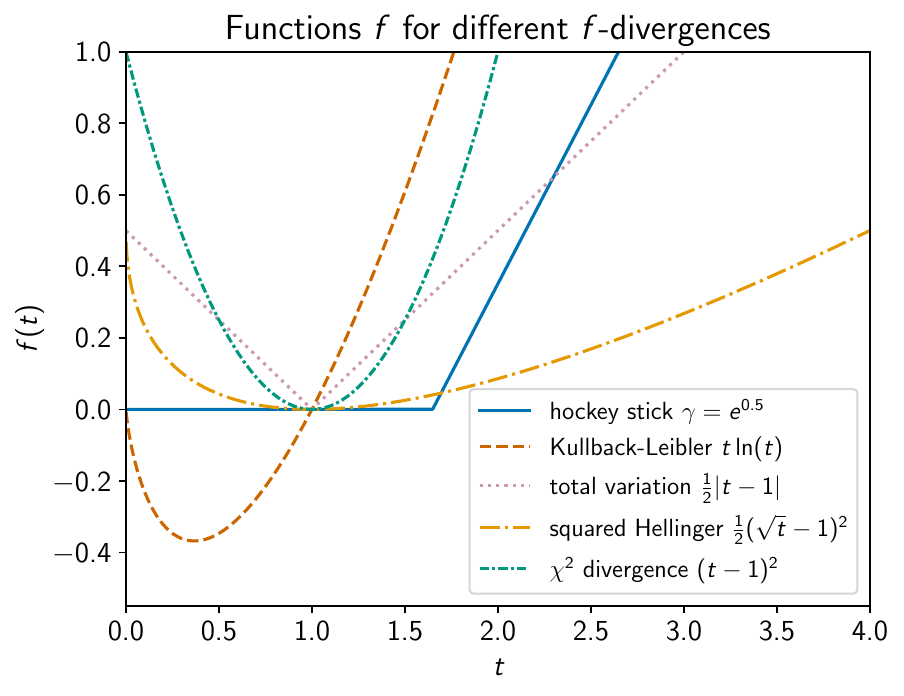}
    \caption{Illustration of the functions $f$ for different $f$-divergences: hockey-stick, Kullback-Leibler, total variation distance, squared Hellinger, and $\chi^2$.}
    \label{fig:divergences}
\end{figure}

For a convex function $f \colon [0,\infty) \to (-\infty,\infty]$ with $f(1) = 0$, the $f$-divergence between two distributions $P$ and $Q$ is
    \begin{align}
    \fdiv{f}{P}{Q} = \int_{\aout} f\parens*{ \frac{dP}{dQ} } dQ.
    \label{eq:fdiv}
    \end{align}
In the first definition of the PLRV in \eqref{eq:plrv}, we can identify $P_{Y|1}$ with $\mc{D}$ and $P_{Y|0}$ with $\mc{D}'$ for a neighboring pair $(\mc{D},\mc{D}')$. By  setting $P = P_{Y|\mc{D}}$ and $Q = P_{Y|\mc{D}'}$ in \eqref{eq:fdiv}, we can easily write any $f$-divergence in terms of the PLRV:
\begin{equation}\label{eq:fdiv-via-PLRV}
    D_f(P \| Q)=\mbb{E}_{Q}[f(e^L)]=\mbb{E}_{P} [e^{-L}\,f(e^L)].
\end{equation}
Since almost any useful divergence is either an $f$-divergence or a simple function of an $f$-divergence (e.g., R\'enyi divergence), this demonstrates the universality of the PLRV. Figure \ref{fig:divergences} shows the function $f(\cdot)$ corresponding to some common divergences.

For the special case of $f(t) = (t - \gamma)^+ + (1 - \gamma)^+$, we obtain what is called the $\Egam$ or ``hockey-stick'' divergence~\cite{BartheOlmedo:13divergence,polyanskiy2010channel} written as
    \begin{align}
    \Ediv{P}{Q} &=\mbb{E}_P[(1-\gamma e^{-L})^+]+(1-\gamma)^+
    \\&=    \mathbb{P}_{P}\parens*{ L > \log \gamma } 
        - \gamma \mathbb{P}_{Q}\parens*{ L > \log \gamma } 
        \notag \\
        &\hspace{4cm} 
            +(1 - \gamma)^+
    \label{eq:Ediv}
    \end{align}
where the first form follows from \eqref{eq:fdiv-via-PLRV}, and the second follows from some simple manipulations from the definition of $f$-divergence in \eqref{eq:fdiv}. We can see that the hockey-stick divergence directly gives us $(\dpeps,\dpdel)$-DP by setting $\gamma=e^\dpeps$. That is, 
\begin{equation}
    \mathsf{E}_{e^\dpeps}(P\mid\mid Q)=\mbb{P}_P(L>\dpeps)-e^\dpeps \mbb{P}_Q(L>\dpeps).
\end{equation}
Comparing to \eqref{eq:def:dp} we see that given $\dpeps$, $\mathsf{E}_{e^\dpeps}$ is nothing but $\dpdel$! For a family of distributions $\{ P_{Y|\mc{D}} : \mc{D} \in \mc{X}^n \}$ the parameters $(\dpeps,\dpdel)$ in the DP definition are related by
    \begin{align}
    \dpdel(\dpeps) 
        &= \sup_{\mc{D} \sim \mc{D}'}
        \mathsf{E}_{e^\dpeps}(P_{Y|\mc{D}}\mid\mid P_{Y|\mc{D}'})
        \notag \\
        &= \sup_{\mc{D} \sim \mc{D}'} 
        \expect*{\left(1 - e^{\dpeps - \plrv_{\mc{D}, \mc{D}'}} \right)^{+} }.
    \label{eq:worstdelta:hsd}
    \end{align}
This shows that understanding more about $f$-divergences in general and the hockey-stick divergence in particular can give us insight into the guarantees differential privacy is making. In the information theory literature, the connection between $E_\gamma$ divergence and hypothesis testing (or, in the DP case, $f$-DP) has been leveraged to derive finite-blocklength channel capacity bounds \cite{polyanskiy2010channel}.  

\section{Composition and privacy accounting}


A core challenge in differential privacy is \emph{privacy accounting}: tracking the value of the  $(\dpeps,\dpdel)$ parameters when data is queried multiple times. Precise accounting is critical for privacy-preserving machine learning, where privacy-sensitive data is queried thousands of times during model training. Each query incurs a small privacy loss, which accumulates over time. An important distinction in composition analyses is whether the queries are adaptive (dependent on previous queries) or non-adaptive. Stochastic optimization methods for differentially private machine learning are adaptive: the query at the current iteration is a function of query responses at past iterations. Here we focus on the simpler case of non-adaptive composition in which we can understand the main ideas and approaches.

\subsection{Composing Privacy Mechanisms}
\label{sec:composition}

The expression in  \eqref{eq:worstdelta:hsd} suggests that analyzing the PLD can help in finding the best $(\dpeps,\dpdel)$ tradeoff for a given mechanism. This is especially useful in analyzing what happens when multiple mechanisms are applied to the same private data. If we release random variables $\rout_1, \rout_2, \ldots, \rout_k$ from $k$ mechanisms $P_{\rout_1|\mc{D}}, P_{\rout_2|\mc{D}}, \ldots, P_{\rout_k|\mc{D}}$ applied to the same dataset, the corresponding PLRV for a pair of neighbors is a sum:
    \begin{align}
    \plrv_{\mc{D},\mc{D}'} 
    &= \log \prod_{j=1}^{k} \frac{dP_{\rout_j|\mc{D}}}{dP_{\rout_j|\mc{D}'}}( \rout ) 
    = \sum_{j=1}^{k} \plrv_{j,\mc{D},\mc{D}'},
    \label{eq:plrv:sum}
    \end{align}
where $\plrv_{j,\mc{D},\mc{D}'}$ is the PLRV for the $j$-th mechanism. In the differential privacy literature, this is known as the problem of \emph{composition} and there have been many proposed approaches to understand the the total PLRV $\plrv_{\mc{D},\mc{D}'}$ based on properties of the individual PLRVs $\plrv_{j,\mc{D},\mc{D}'}$.
If these mechanisms are independent then analyzing the PLRV is equivalent to analyzing the distribution of a sum of random variables: this is a fundamental question in probability theory for which we have a variety of approaches.

\subsection{Basic Composition}

The simplest approach, known as \emph{basic composition}, only uses information about the support of the PLRV. For example, if each individual channel is $(\dpeps,0)$-DP, the PLRV is almost surely upper bounded by $e^{\dpeps}$, so the PLD has finite support. This implies that the PLD from $k$ mechanisms will also have bounded support and the PLRV will be upper bounded by $e^{k\dpeps}$, so the overall mechanism is $(k\dpeps,0)$-DP. A similar argument shows that composition of $k$ mechanisms that are $(\dpeps,\dpdel)$-DP will guarantee $(k\dpeps,k\dpdel)$-DP. This approach uses only the simplest information about the individual PLDs: if we only know $(\dpeps_j,\dpdel_j)$ values for the individual mechanisms there is an exact formula for the overall $(\dpeps,\dpdel)$ guarantee but computing it can be $\#P$-complete~\cite{KairouzOV:17composition,MurtaghV:16complexity}.

\subsection{PLD-Based Accounting}

If we know more about the actual channels used in the privacy mechanisms, there are three broad classes of approaches for understanding the sum in \eqref{eq:plrv:sum}. Since the PLD of the sum is the convolution of the individual PLDs, a natural numerical approach for calculating the overall privacy loss would to use characteristic functions~\cite{zhu2022optimal} and the FFT to take the Fourier transforms of the (approximate) PLDs. For a more exact approach, we can instead use the moment generating or cumulant generating function to prove measure concentration results about the PLD. Finally, we can use approaches based on the central limit theorem to understand the overall privacy loss. Each of these approaches to calculating or bounding the PLD provides what the privacy literature calls a \emph{privacy accountant}.

\subsubsection{FFT-based accounting} If the PLD of each mechanism is known then, for a fixed pair of datasets $\mathcal{D}$ and $\mathcal{D}'$, the PLD of the sum in \eqref{eq:plrv:sum} is the  convolution of the individual PLDs. This convolution can be accurately approximated numerically using the FFT~\cite{koskela2020computing}, and the resulting PLD used to estimate the expectation in \eqref{eq:worstdelta:hsd}. 

The main challenge in applying this FFT-based approach is resolving the supremum in \eqref{eq:worstdelta:hsd}. In fact, the ``worst-case'' neighboring datasets $\mathcal{D}\sim \mathcal{D}'$  may change with the number of compositions! This challenge can be circumvented by identifying a \emph{dominating pair} of distributions. 

A pair of distributions $(P,Q)$ is said to be a \emph{dominating pair} for a privacy mechanism $P_{Y|\mathcal{D}}$ if the hypothesis testing error tradeoff curve $\Pmd(\Pfa)$ for the pair $(P,Q)$ lies below that of any pair $(P_{Y|\mc{D}},P_{Y|\mc{D}'})$ for $\mc{D} \sim \mc{D}'$. In terms of the hockey-stick divergence this means that for every $\dpeps$ we have
\begin{equation*}
 \sup_{\mc{D} \sim \mc{D}'}
        \mathsf{E}_{e^\dpeps}(P_{Y|\mc{D}}\mid\mid P_{Y|\mc{D}'}) \leq  \mathsf{E}_{e^\dpeps}(P\mid\mid Q).
\end{equation*}
By using the dominating pair $(P,Q)$, we can apply FFT-based accounting without identifying the specific worst-case dataset pair that achieves the supremum in \eqref{eq:worstdelta:hsd}. Instead, the dominating pair $(P,Q)$ defines a ``dominating'' PLRV:
    \begin{align}
        L=\log \frac{dP}{dQ}(Y), \qquad Y\sim  P.
    \end{align}
While this approach may yield a slightly looser bound on $\dpdel(\dpeps)$, it significantly simplifies privacy accounting. Several FFT-based accounting methods are available, many with open-source implementations~\cite{koskela2020computing,koskela21aFFT,AlghamdiACKS:24isit,SPA,saddlepoint:code,gopi2019numerical,ConnectingDots,FasterPrivacyAccountant,zhu2022optimal}.

\subsubsection{Tail bounds on the PLRV} The earliest work on improving bounds for composition used measure concentration on the sum in \eqref{eq:plrv:sum}. Because exact composition guarantees~\cite{KairouzOV:17composition} are difficult to compute~\cite{MurtaghV:16complexity}, many alternative definitions of privacy have been proposed that provide more refined bounds on the PLRV, such as zero-concentrated DP~\cite{BunS:16zcdp,BunDRS:18truncated} and R\'{e}nyi DP~\cite{Mironov:17renyi,WangBK:19analytical}. In fact, R\'{e}nyi DP amounts to keeping track of the cumulant generating function of the PLRV, which can be converted to tail bounds (and $(\dpeps,\dpdel)$ guarantees) via standard concentration inequalities.


\subsubsection{CLT-based approaches} 
The FFT approaches exploit the fact that the density of a sum of random variables is the convolution of the densities to numerically approximate the moment generating functions of the distributions. By working with the moment or cumulant generating functions directly, we can use the central limit theorem to understand the asymptotic behavior of composed mechanisms~\cite{SommerMM:19clt,SPA,Cactus,Isotropic,DongRS:21gaussiandp,zhu2022optimal,CannoneKS:20discrete}. A recent strategy that provides precise analytical bounds for privacy accounting is the \emph{saddle-point accountant} \cite{SPA} which, broadly speaking, first considers the distribution of a tilted PLRV, then applies the CLT to approximate privacy guarantees under composition.







\section{Optimization of privacy mechanisms}

The preceding discussion has focused entirely on the privacy guarantees of mechanisms/channels but not on the impact of these channels on the usefulness, or \emph{utility} of the mechanism output. The privacy of a DP channel is a function of the channel alone, whereas the utility will in general depend on both the channel and an assumed distribution on $\mc{X}^n$ for the data set $\mc{D}$. This raises the natural question of \emph{channel optimization}: among all channels/mechanisms which guarantee the same privacy parameters $(\dpeps,\dpdel)$, which ones will have the highest expected utility?  A very important point to note is that the privacy properties of the channel must hold for any distribution on $\mc{D}$, whereas the utility is measured with respect to the randomness in $\mc{D}$ and the privacy channel. 

Many information theory problems involve optimization over a set of distributions, so the idea of optimizing over channels is familiar territory for information theorists.
For example, the rate-distortion function is calculated by optimizing a channel (i.e., a conditional distribution) to minimize the mutual information subject to a distortion constraint. As we will see, the rate-distortion optimization problem is closely related to optimization of privacy mechanisms. In the rate-distortion problem, the channel from the source to the reconstruction\footnote{Not to be confused with the \emph{test channel}, a term commonly used for the reverse: from the reconstruction back to the source. Here we focus only on the forward channel.} can be viewed as a privacy mechanism. The main difference is that instead of optimizing a mutual information, we take a privacy risk measure as the objective function to minimize: this gives the problem and solution a very different flavor.


\subsection{Queries and Sensitivity}

Setting up an optimization problem for privacy mechanisms requires some assumptions on the structure of the mechanism. Without any such assumptions, we do not know what spaces the dataset and outputs live in, or what information is desired from this dataset. Unlike rate-distortion theory, DP assumes no prior distribution on the dataset; it instead relies only on the notion that we compare two neighboring datasets. If we're not careful in setting up our desired specifications, it can be impossible to find \emph{any} privacy mechanism with meaningful privacy and utility values. Indeed, the following example, perhaps the most straightforward imaginable setting where privacy might be desired, turns out to be fundamentally infeasible.

Suppose our dataset $\mc{D}=(x_1,x_2,\ldots,x_n)$ consists of scalars $x_i\in\mathbb{R}$, and the goal is to find the average
\begin{equation}
    \bar{x}=\frac{1}{n} \sum_{i=1}^n x_i.
\end{equation}
We want to find a mechanism $P_{\rout | \mc{D} }$ that achieves $(\dpeps,\dpdel)$-DP, and a worst-case MMSE utility constraint:
\begin{equation}
    \mathbb{E}[(Y-\bar{x})^2|\mc{D}]\le \sigma^2\text{ for all }\mc{D}.
\end{equation}
This turns out to be impossible for \emph{any} non-trivial\footnote{That is, $\dpeps$ and $\sigma$ finite, and $\dpdel<1$.} values of $\dpeps$, $\dpdel$, and $\sigma$. The reason is that for neighboring $(\mc{D},\mc{D}')$, the associated averages $\bar{x}$ and $\bar{x}'$ could be arbitrarily far apart, since the $\{x_i\}$ are unbounded. Thus, in order to satisfy the utility constraint, the conditional distributions $P_{\rout|\mc{D}}$ and $P_{\rout|\mc{D}'}$ need to be largely separated, which will violate the DP requirement.

The above example can be saved by imposing a prior distribution on the entries of the dataset---precisely what we do in the rate-distortion problem. However, if we can design the privacy mechanism with knowledge of this distribution, then the privacy problem itself becomes moot, since all we wanted was the expectation in the first place. For instance, if I know the distribution of hair loss in cancer patients, then I don't need to collect any data from actual cancer patients to find the average amount of hair loss!

A better solution to this fundamental limitation is to make a weaker assumption on the dataset: in the above example, if we assume instead that that the entries of the dataset are bounded, then the difference between the averages for neighboring datasets $\bar{x}$ and $\bar{x}'$ is also bounded. This amounts to a requirement to find a mechanism that is universal among all distributions for the data with bounded support. We can generalize this idea by assuming we have a \emph{query function} $q \colon \mc{X}^n \to \mathbb{R}^d$ whose values are close on neighboring datasets. We say that $q(\cdot)$ has \emph{sensitivity} $s$ with respect to a norm $\norm{\cdot}$ on $\mathbb{R}^d$ if 
\begin{equation}
    s = \max_{\mc{D}\sim\mc{D'}} \norm{ q(\mc{D})-q(\mc{D'}) }.
\end{equation}
Given a query function with a finite sensitivity, the privacy mechanism typically has two parts: first evaluate the query function, and then put the query value through a random channel which is designed to have the desired privacy and utility characteristics. Now we can optimize for this privacy channel independent of the query function. In fact, we only need to keep track of the sensitivity $s$\footnote{This assumes the utility and sensitivity are using the same norm.} and can forget about all other details of the query function and the application at hand.

\subsection{Optimization Problems for Privacy Noise}

In this section we will focus on scalar query functions ($d = 1$) to allow for easier visualization: extensions to higher dimensions can be done by adding per-coordinate noise.
The earliest approach to finding a privacy mechanism via optimization in this vein is by Geng et al.~\cite{GengKOV:15staircase}, who considered a very simple setting: a mechanism $Y=q(\mc{D})+Z$ for guaranteeing $\dpeps$-DP (meaning that $\dpdel$ is assumed to be $0$), where $Z$ is additive noise independent of all other variables.

\begin{figure}[htb]
    \centering
    \includegraphics[width=0.7\columnwidth]{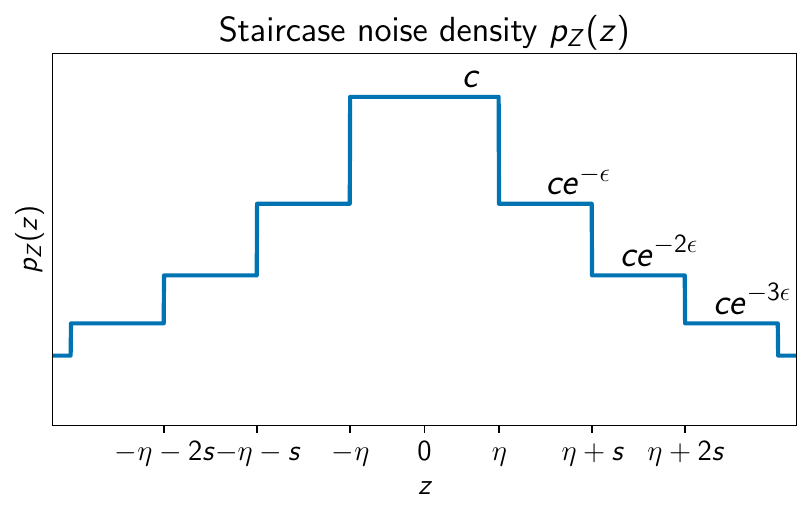}
    \caption{The staircase distribution, which optimizes $\dpeps$-DP for a single composition with sensitivity $s$.}
    \label{fig:staircase}
\end{figure}

Let $\Delta(\dpeps)$ be the set of all distributions for $Z$ that guarantee $\dpeps$-DP for a query $q(\cdot)$ of sensitivity $s$. The resulting optimization over the distribution $p_{z}$ of $Z$ is 
\begin{equation}\label{staircase-opt}
\begin{array}{ll}
\displaystyle\minimize_{p_Z(z) \in \Delta(\dpeps)} & \displaystyle \max_{|a|\le s} \ \max_{z} \ \log \frac{p_Z(z)}{p_Z(z-a)},\\
~\\
\text{subject to } &\mathbb{E}[c(Z)]\le C.
\end{array}
\end{equation}
The constraint $\mathbb{E}[c(Z)]\le C$ ensures that the noise has a limited expected cost with respect to a cost function $c$ (typically quadratic, but could be anything) that captures the utility of producing the noisy $Y = q(\mc{D})+Z$ when the true query value is $q(\mc{D})$. The objective function represents the maximum $\dpeps$-DP for a given sensitivity $s$: the term $\log p(z)/p(z-a)$ represents the PLRV for two query function values that differ by $a$, so we need to maximize over all $|a|\le s$. For a well-behaved cost function $c$, this problem has a closed-form solution known as the \emph{staircase mechanism}, illustrated in Figure~\ref{fig:staircase}, with PDF given by
\begin{equation}
p_Z(z)=c\, e^{-k\dpeps},  |z|\in[(k-1)s+\eta, k\,s+\eta],\ k\ge 0
\end{equation}
for suitably chosen constants $c,\eta,\dpeps$ ($\dpeps$ is the resulting DP parameter).

Once we consider the composition of many mechanisms, the staircase mechanism is no longer optimal. A natural limit to take from an information theory perspective is the \emph{large composition regime}---that is, where the number of compositions goes to infinity. In particular, consider the decomposition of the PLRV in \eqref{eq:plrv:sum} where the number of compositions $k\to\infty$. Since we can take the individual PLRVs $L_{j,\mathcal{D},\mathcal{D}'}$ to be independent, by the law of large numbers the total PLRV $L_{\mc{D},\mc{D'}}$ is close to its expectation. As we saw in \eqref{kl-divergence}, the expectation of the PLRV is simply the KL-divergence between the two distributions. As seen by the expression for $\dpdel(\dpeps)$ in \eqref{eq:worstdelta:hsd}, we will generally achieve better privacy characteristics if the PLRV is smaller. Thus, the optimal mechanism for the large composition regime will be the one that minimizes the KL-divergence. 

Alghamdi et al.~\cite{Cactus} determined the optimal additive noise for a large number of compositions by solving the following optimization problem:
    \begin{equation}
    \begin{array}{ll}
    \displaystyle\minimize_{p(z)} & \displaystyle \max_{|a|\le s} \ \kldiv{p(z)}{p(z-a)},\\
    \text{subject to} & \mathbb{E}[c(Z)]\le C.
    \end{array}\label{cactus-opt}
    \end{equation}
As in the optimization problem in \eqref{staircase-opt}, we must maximize over all possible shifts $a$ up to the sensitivity, and we impose an expected cost constraint on the noise distribution. Due to this optimization representing the fundamental limit in the large composition regime, the optimal value in \eqref{cactus-opt} can be thought of as \emph{privacy capacity}~\cite{AlghamdiACKS:24isit}.

\begin{figure}[htb]
    \centering
    \includegraphics[width=0.7\columnwidth]{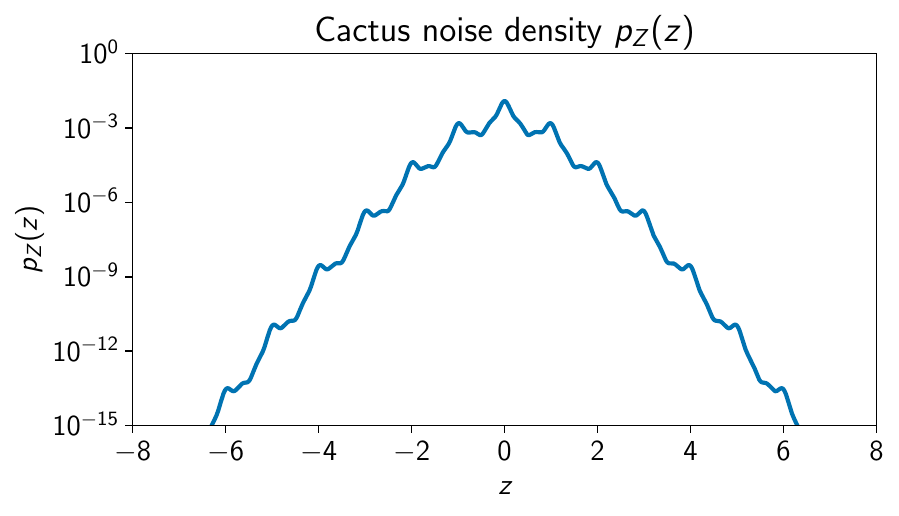}
    \caption{The noise distribution for a Cactus mechanism, plotted on a semi-log scale. The cost constraint is $\expect{Z^2}\le 0.25$, and the sensitivity is $s=1$.}
    \label{fig:cactus}
\end{figure}

Despite the close resemblance between the optimization problem in \eqref{cactus-opt} and the rate-distortion optimization problem, they produce significantly different results. For quadratic cost, the optimal rate-distortion channel would be the Gaussian noise channel, but for this problem, the result is a non-monotonic, ``spiky'' distribution shown in Figure \ref{fig:cactus}, dubbed the \emph{Cactus mechanism}~\cite{Cactus}. The optimization problem has no closed-form solution: the distribution in Figure \ref{fig:cactus} comes from choosing a finite-dimensional but large space of possible distributions and solving the problem exactly within this space. This approach yields distributions that are provably close to optimal. In follow-up work~\cite{Isotropic}, a similar approach was taken for multi-dimensional noise.

The above discussion may suggest that using Gaussian noise---by far the most popular approach in practice for $(\dpeps,\dpdel)$-DP---is a bad idea, since it is strictly suboptimal compared to the Cactus. While this is technically true, we can look at the problem differently to see that Gaussian noise is good, actually. In particular, taking the optimization problem in \eqref{cactus-opt} to the \emph{small sensitivity regime}, meaning $s\to 0$, then we can again find closed-form solutions for the optimal noise. In \cite{Fisher}, it was shown that the optimal noise in this regime will minimize the Fisher information, which in turn is the noise which solves a differential equation equivalent to \emph{Schr\"odinger's equation}. In particular, for cost function $c$, the optimal noise in the small-sensitivity regime has PDF $p(z)=y(z)^2$, where $y$ is the solution to
\begin{equation}
    y''(z) = (\theta\,c(z) - E)y(z),
\end{equation}
where $\theta$ and $E$ are constants to be selected so that the resulting noise distribution satisfies the cost and normalization constraints. Mechanisms derived from this approach are called \emph{Schr\"odinger mechanisms}.
If we take $c(\cdot)$ to be the quadratic cost, the optimal distribution in the Schr\"odinger mechanisms is in fact the Gaussian, which explains why using Gaussian noise often works well in practice. However, for other cost functions, the optimal noise distribution can be quite different. For example, for expected absolute value cost, the optimal noise distribution is derived from the Airy function~\cite{Fisher}.








\section{The view ahead}


Predictive and generative machine learning models are rapidly making inroads into various human decision-making algorithms. These models are often trained using private or sensitive data, sometimes without the explicit consent of the individuals involved. Developing the tools to measure how much released models reveal information and how to protect that information is an ethical imperative. Differential privacy has emerged as a strong contender for providing the kinds of rigorous privacy guarantees that are needed. However, there are still many challenges for the future. We close this article by highlighting a few of these.

\subsection{Managing privacy in training large ML models}

As ML models grow in size (current ``small'' foundation models have at least 1B parameters), so does the training time, and therefore the number of iterations required to learn models with algorithms such as DP-SGD \cite{ACGMMTZ16}. Many of the advances in privacy accounting have been aimed at developing a better analysis of existing algorithms and in particular DP-SGD. To apply  composition approaches such as the saddle-point accountant \cite{SPA} or the connect-the-dots algorithm \cite{ConnectingDots} requires analyzing the algorithm that is actually implemented since small changes can impact the formal privacy guarantees.

As an example, a key technique in improving the performance of differentially private stochastic optimization algorithms is to use random subsampling, which often amplifies privacy~\cite{KLNRS11,LiQS:12sampling,KairouzOV:17composition,BalleBG:2020sampling}. For training machine learning models, the mini-batches to use in stochastic gradient descent can take advantage of this amplification~\cite{ACGMMTZ16} by using Poisson subsampling. Despite its theoretical appeal, Poisson subsampling is rarely used in practice and many implementations randomly shuffle the data while performing a privacy analysis using Poisson subsampling~\cite{ponomareva2023dp}.~\footnote{The Opacus library~\cite{opacus} does implement Poisson subsampling but works only for datasets which can fit into memory.} This can cause challenges for composition methods which rely on the PLRVs for each mechanism to be sampled  independently~\cite{chua2024how,lebeda2024pitfalls,chua2024scalable}. While an improved analysis for the case of gradient algorithms is not likely possible~\cite{chua2024how}, understanding how to handle dependence between PLRVs more generally is an interesting challenge for future research.


\subsection{Privacy while fine-tuning ML models}

Concerns about the computational complexity, sustainability, and costs of training large models have led to a growing interest in reusing and fine-tuning publicly available foundation models for user-specific tasks. For example, a small hospital or research unit with limited computational budget may want to fine-tune a publicly available vision transformer (ViT) model for lung cancer X-ray classification. Although these foundation models may, in general, not be trained privately, assuring privacy of the dataset used to fine-tune may be imperative in many settings where the additional data is private or sensitive. The problem of assuring privacy using fine-tuning methods such as low-rank adaptive tuning or simple last layer retraining, which significantly reduce the training parameter size, hold some promise and need to be further explored. 

\subsection{Synthetic data generation}

Synthetic data, or data which ``looks like'' a provided private data set, is a compelling approach for protecting privacy in a ``one-shot'' manner. For example, we could use a method to learn a generative model for synthetic data; the learning still requires privacy guarantees. Privacy-preserving synthetic data generation has remained an active area of research. Raskhodnikova et al.~\cite{raskhodnikova2021differentially} discusses theoretical foundations and Hu et al.~\cite{HuWLLGGDFLS:24sok:synthdata} summarize most of the existing approaches. From an information theorist's perspective, both characterizing fundamental limits (converse bounds or impossibility results) and schemes that meet those limits (achievability results or algorithms) are interesting. More research is needed to determine whether or not it is possible to have a DP mechanism that can generate~\emph{multiple} samples from a private distribution. A significant challenge in synthetic data is to develop mechanisms that can guarantee some ``closeness'' between general (not pre-specified) queries on the private and released samples.


\subsection{Selecting privacy parameters in practice} 

How should we select the parameters of a privacy mechanism? This is arguably the most important challenge facing practitioners interested in using differential privacy. For instance, when training a model using DP-SGD, what should be the variance of the noise added to gradients? Typically, the target value of $\dpdel$ is fixed to be $\ll n$ (for many data sets this is between $10^{-8}$ and $10^{-5}$), and  the  parameters of a privacy mechanism---the ``noise'' in the channel---are calibrated to achieve a certain value of $\dpeps$. This leads to an  important practical challenge: \emph{what are ``acceptable'' values of $\dpeps$ and $\dpdel$?} There has been a flurry of recent works that argue that DP guarantees are more easily understood by practitioners when they are converted to operational \emph{attack risks}~\cite{cummings2021need,cummings2024attaxonomy}. For instance, converting $(\dpeps,\dpdel)$ to the hypothesis testing interpretation of DP, introduced in Section \ref{sec:hyp}, is one example of using an operational interpretation of the DP guarantees. Recent work \cite{kulynych2024attack} has demonstrated that calibrating a privacy mechanism to satisfy a given level of attack risk---as opposed to targeting a less-interpretable $(\dpeps,\dpdel)$ value---can significantly increase utility. An interesting future research direction is understanding if it is possible to keep the ``differential'' part of DP, but perhaps moving away from $(\dpeps,\dpdel)$ (or, equivalently, $\mathsf{E}_{e^\dpeps}$-divergence), by focusing instead on more interpretable quantities such as $f$-DP.

\subsection{The end of the tour}

In this article we took a tour of different facets of differential privacy through the lens of information theory. While many of the tools are the same (testing, divergences, measure concentration, optimization), the differences between the operational needs for privacy measures differs from the bread-and-butter applications of information theory such as compression or reliable communication. The operational interpretation of the DP guarantee results in considering different divergences between distributions and different kinds of optimization problems. As we have seen, the resulting solutions, such as optimal additive noise distributions, can be significantly different than those encountered in a first course in information theory. There are likely many more places where ``thinking like an information theorist'' can yield novel insights and new strategies for improving privacy guarantees or broadening our understanding of the general problem of quantifying privacy. We hope that  this brief tour of some highlights will help you, Dear Reader, to tackle some of these future challenges.

\section*{Acknowledgments}

The authors thank Prof.~Shahab Asoodeh for helpful discussions and suggestions. The work of Sarwate was sponsored in part by the US National Institutes of Health under award 2R01DA040487 and in part by the US National Science Foundation under grant CNS-2148104, which is supported in part by funds from federal agency and industry partners as specified in the Resilient \& Intelligent NextG Systems (RINGS) program. The work of Kosut and Sankar is supported in part by the US National Science Foundation grants CIF-1901243 and CIF-2312666. Sankar's work is further supported by NSF grant CIF-2007688. Calmon's work is supported by US National Science Foundation grant CIF-2312667 and CIF-2040880.

\bibliographystyle{IEEEtran}
\bibliography{privacy}

\end{document}